\documentclass{WileyMSP-template}

\usepackage{amsmath,amssymb}
\usepackage{graphicx}
\usepackage{bm}
\usepackage{chemformula}
\usepackage{multirow,makecell}

\begin{document}

\pagestyle{fancy}
\rhead{\includegraphics[width=2.5cm]{Fig.eps}}

\title{Controllable Excitation of Surface Plasmon Polaritons in Graphene-Based Semiconductor Quantum Dot Waveguides}

\maketitle


\author{Mikhail Yu. Gubin}
\author{Alexei V. Prokhorov*}
\author{Valentyn S. Volkov}
\author{Andrey B. Evlyukhin}



\begin{affiliations}
Dr. M. Yu. Gubin, Dr. A. V. Prokhorov\\
Department of Physics and Applied Mathematics, Vladimir State University named after A. G. and N. G. Stoletovs (VlSU)\\
Gorky st. 87, Vladimir, 600000, Russian Federation\\
Email Address: avprokhorov33@mail.ru

Dr. M. Yu. Gubin, Dr. A. V. Prokhorov, Dr. V. S. Volkov\\
Center for Photonics and 2D Materials, Moscow Institute of Physics and Technology (MIPT)\\
Institutskiy per. 9, Dolgoprudny, 141700, Russian Federation

Prof. A. B. Evlyukhin\\
Institute of Quantum Optics, Leibniz Universit\"{a}t Hannover (LUH)\\
Welfengarten 1, Hannover, 30167, Germany

\end{affiliations}



\begin{abstract}

In the present paper, the collective near-field effects in a two-graphene sheets plasmonic waveguide loaded with an array of $\textrm{Ag}_{2}\textrm{Se}$ quantum dots excited by external radiation are theoretically studied. This research aims to develop a theoretical approach to realizing controllable excitation and the propagation of surface plasmon polaritons (SPPs) in planar graphene waveguides. The proposed model is based on the semiclassical description that implies the dispersion equation's solution for SPPs excited in two coupled graphene sheets and the embedded quantum dots' energy spectra analysis. Using the finite difference time domain method, the different near-field patterns realized in the waveguides depending on the quantum dots' ordering and an external voltage applied independently to each graphene sheet with additional gold electrodes are numerically investigated. The research shows that applying external voltages allows locally changing the graphene chemical potential, thereby leading to controllable excitation of SPPs in the space between the electrodes. Under these conditions, the propagation direction of the excited SPPs is determined by the geometrical configurations of the gold electrodes providing the required SPP routing. The investigated system offers new opportunities for near-field energy transport and concentration, which can be applied to develop ultra-compact photonic devices.

\end{abstract}


\section{Introduction}
The realization of light control on the nanoscale is a fundamental problem of modern nano-photonics and plasmonics~\cite{boj1}. In this context, all-graphene-based technologies seem to be suitable approaches for solving the problem since the electromagnetic field is well-localized on graphene~\cite{ali,Grigorenko}, whereas the quantum dots (QDs) can be highly efficient converters of light into the near-field~\cite{Koppens} in the wide spectral range. At the same time, the questions of the experimental development of graphene systems with the required properties and the creation of appropriate interfaces have been opened so far.

When the circuit's speed is not a priority, simple voltage control can be used to manipulate the localized field states on graphene~\cite{n02,n2,n3,n4,n5,n6}. In its pristine condition, graphene has a vanishing density of states at the Fermi level, which is not good for superconductivity still, if it can be heavily electron-doped, it is predicted to become a superconductor~\cite{Gonzalez}. The conventional commercial graphene has the chemical potential $\mu_{c}$ of about $0.1 \; \textrm{eV}$, and for doped graphene, it can reach values of up to $0.6 \; \textrm{eV}$~\cite{ali}, but these values are fixed. The main effect for controlling the chemical potential is the dependence of the graphene carrier density on the external voltage (or electric field strength)~\cite{Gomez,Zheng}. Note that the modifications of the chemical potential lead to a change in the excitation and the propagation characteristics of surface plasmon polaritons (SPPs) supported by graphene layers~\cite{ali}.

The excitation of SPPs on graphene requires the use of either time-tested technologies~\cite{n7} or more lab-consuming approaches that promise much higher light-SPP conversion coefficients. In the latter case, this approach can be associated with the use of classical optical resonances~\cite{n8,Evlyukhin1,Evlyukhin_B} and new quantum-configuration resonances~\cite{Koppens,ourLPR} with nanostructures and quantum dots (QDs) located near the interface of two materials. The remarkable feature of QDs is the ability to tune their photoluminescence frequencies to the desired spectral range by changing their shape and geometrical sizes~\cite{n9}. At the same time, such a point emitter becomes an excellent source of near-field excitations when a suitable distance to the surface is chosen~\cite{ourASS,Chance}. However, for efficient excitation of plasmon polaritons in graphene (under condition $\hbar \omega<2\mu_{c}$, where $\hbar \omega$ is the photon energy of the pump field), QD photoluminescence in the mid-wave infrared (MWIR, $3--6 \; \textrm{\textmu m}$) range is required, which significantly reduces the choice of semiconductors for their fabrication. Suitable semiconductor QDs must have a very small band gap for emitting such wavelengths on an interband transition or large values of scattering cross-sections on intraband transitions for the first levels of the conduction band~\cite{Shesterikov}. In this sense, good candidates for SPP generation on planar or multilayer graphene sheets are HgTe QDs with bright photoluminescence in MWIR emitted by both intra- and interband transitions~\cite{Tang,Goubet}. Other candidates are less toxic and more stable $\textrm{Ag}_{2}\textrm{Se}$ QDs~\cite{Sahu,Park}, which produce power photoluminescence at a wavelength near $5 \; \textrm{\textmu m}$.

Note that single QDs or their disordered arrays cannot provide sufficient efficiency of light-SPP conversion. Instead, cooperative phenomena and collective near-field effects~\cite{EVL4} with highly localized states of the electromagnetic field can be used. Such effects are self-phasing in the form of superradiance~\cite{n11} and localization due to defects~\cite{n12}.

In the present paper, we focus on investigating optimal configurations of $\textrm{Ag}_{2}\textrm{Se}$ QD arrays loaded into a graphene waveguide~\cite{Kim2} to achieve a strong near-field due to QDs excitation by an external pumping wave. We also numerically study the possibility to control the subwavelength light energy concentration in graphene systems and advanced two-dimensional materials~\cite{n13} with QD structures by applying an external voltage. Such a combination of ordered nanostructures with an external modulation of their physical characteristics~\cite{Garanovich} (provided, for example, by an electric voltage) allows extending the ways to control the localization of light at the nanoscale. Similar approaches are already extensively used in plasmonics and nanophotonics to create bright light sources~\cite{Winkler}, dense coding of information~\cite{Carvalho}, and velocity light control~\cite{Wang3}. Such systems can also include modern topological lasers based on the combination of a pair of resonant structures with considerably different geometries and different scales, which leads to the creation of high concentrations of optical energy~\cite{Bahari,Bandres}.

In our model, we consider a combination of two systems. The first is a graphene waveguide loaded with quantum dots. The second is a capacitor consisting of a gold stripe and a graphene sheet of the waveguide. The capacitor system is used for the realization of the local voltage control of graphene chemical potential $\mu_{c}$. We investigate the possibility of satisfying the condition $\hbar\omega<2\mu_{c}$ for the SPP excitation due to the relaxation of the excited QDs by applying an external voltage. Because the voltage will act locally, just below the gold stripe, the graphene SPPs can be excited only in this area. The QDs placed outside the area with $\hbar\omega<2\mu_{c}$ are not coupled with the SPPs and, therefore, their relaxation occurs into light and phonons. The suggested system can provide controllable selective excitation and routing of the graphene SPPs.

In our investigation, we use a semiclassical approach to define the graphene waveguide system's main parameters with an $\textrm{Ag}_{2}\textrm{Se}$ QD array that can lead to the excitation of graphene SPPs only in the local space between the gold stripes used to connect external electrical voltages. Using numerical simulation based on the finite difference time domain method, we reveal an unusual regime of a low-frequency modulation of the SPPs excited in the graphene waveguide by the SPPs excited on the gold stripes. Such structures with an SPP-SPP modulation can be suggested for the development of ultra-thin laser systems and all-graphene devices for controlling light at the nanoscale.

\begin{figure}[t]
\centering
\includegraphics[width=0.7\columnwidth]{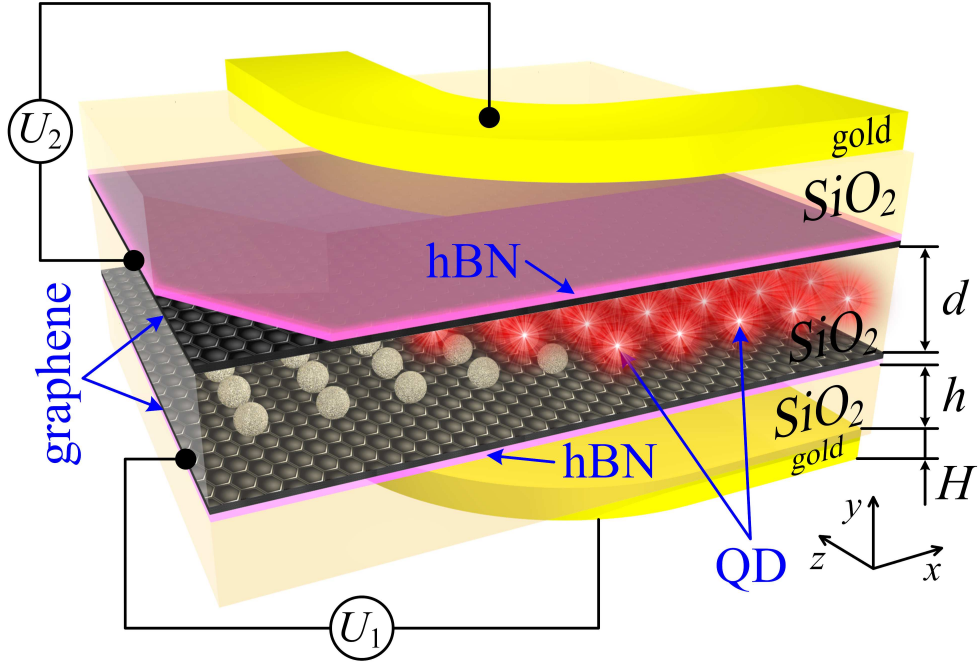}
\caption{\label{fig:1} Layout of the plasmonic waveguide with voltage control $U_{1,2}$ for different graphene sheets, where $h=50 \; \textrm{nm}$, $H=20 \; \textrm{nm}$, $d=14 \; \textrm{nm}$. The size (diameter) of QD is $5.794 \; \textrm{nm}$.}
\end{figure}

\section{Model of SPP generation and voltage control in the two graphene sheets plasmonic waveguide loaded with Ag2Se quantum dots}
The functional block for localization and routing of the near-field energy as a part of the general SPP circuit is shown in Figure~\ref{fig:1}. The graphene plasmonic waveguide composed of two graphene sheets is embedded in the silica host medium. Each graphene sheet is located on a buffer hBN layer, and the gold stripes are deposited to the outer part of the circuit. Then, it is possible to apply a voltage between the selected stripes and the graphene to modify the graphene area's chemical potential near the live stripe.

To characterize and demonstrate the basic functional features of the device presented in Figure~\ref{fig:1}, as a beginning, we consider several important elements of the system.

\begin{figure}[t]
\centering
\includegraphics[width=\columnwidth]{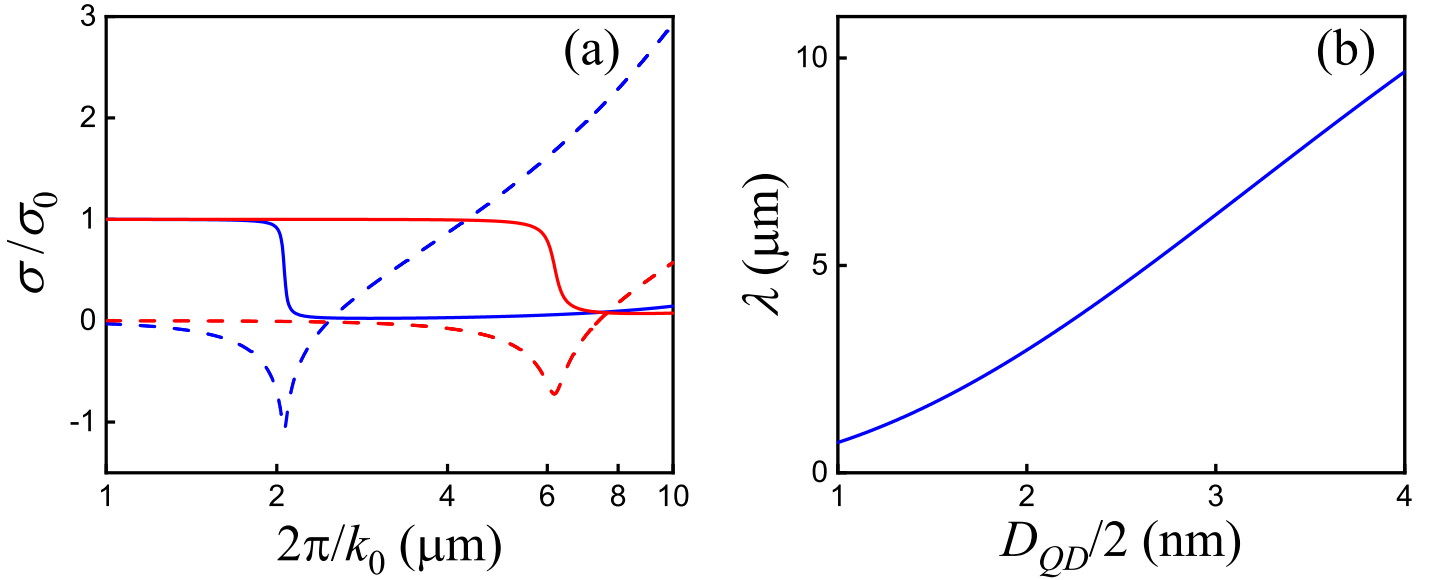}
\caption{\label{fig:2} (a) Real (solid curves) and imaginary (dashed curves) parts of the total $\sigma=\sigma^{\textrm{R}}+i\sigma^{\textrm{I}}$ electric conductivity of graphene normalized to $\sigma_{0}$ versus free-space wavelength for chemical potential $\mu_{c1}=0.1 \; \textrm{eV}$ (red curves), $\mu_{c2}=0.3 \; \textrm{eV}$ (blue curves). (b) Dimensional dependence for the wavelength of the interband transition of $\textrm{Ag}_{2}\textrm{Se}$ QDs with effective masses of the electron $m_{c}=0.32m_{0}$ and the hole $m_{h}=0.54m_{0}$, and $\varepsilon_{st}=11$.}
\end{figure}

The graphene permittivity at the angular frequency $\omega$ can be represented in the form:
\begin{equation}
\label{eq:1}
\varepsilon_{gr}=1+i\frac{\sigma}{\omega \Delta_{g}\varepsilon_{0}},
\end{equation}
where $\Delta_{g}$ is the thickness of the graphene in the surrounding medium with permittivity $\varepsilon_{d}$, $\varepsilon_{0}$ is the vacuum dielectric constant. The graphene conductivity can be described by the Kubo formula~\cite{albert}:
\begin{align}
\label{eq:2}
\nonumber
\sigma\left(\omega,\mu_{c},\tau,T\right)=&\frac{-ie^2/\pi\hbar^2}{\omega +i/\tau } \int^{\infty}_{0}{\epsilon \left(\frac{\partial f_{d}\left(\epsilon \right)}{\partial \epsilon}-\frac{\partial f_{d}\left(-\epsilon \right)}{\partial \epsilon }\right)d\epsilon}\\
&-ie^2\left(\omega +i/\tau \right)/\pi \hbar^2 \int^{\infty}_{0}{\frac{f_{d}\left(\epsilon \right)-f_{d}\left(-\epsilon \right)}{{\left(\omega +i/\tau \right)}^2-4{\left(\epsilon /\hbar \right)}^2}d\epsilon},
\end{align}
where $1/\tau$ is the electron scattering rate, $f_{d}\left(\epsilon \right)=1/\left(e^{\left(\epsilon -\mu_{c}\right)/kT}+1\right)$ is the Fermi-Dirac function, $T$ is the temperature, $k$ is the Boltzmann constant, $\hbar \equiv \frac{h}{2\pi}$ is the reduced Planck's constant, $e$ is the electron charge, $c$ is the speed of light in a vacuum. The first integral for the intraband $\sigma_{\textrm{intra}}$ and the second for the interband $\sigma_{\textrm{inter}}$ conductivity in (\ref{eq:2}) can be approximated in conditions $kT<<|\mu_{c}|,\hbar\omega$ as
\begin{subequations}
\label{eq:3}
\begin{align}
\sigma_{\textrm{intra}}\left(\omega,\mu_{c},\tau,T\right)&=i\frac{8\sigma_{0}kT/h}{\omega +i/\tau}\left(\frac{\mu_{c}}{kT}+2\textrm{ln}\left(e^{-\frac{\mu_{c}}{kT}}+1\right)\right),\\
\label{eq:3}
\sigma_{\textrm{inter}}\left(\omega,\mu_{c},\tau,T\right)&\approx i\frac{\sigma_{0}}{\pi} \textrm{ln}\left(\frac{2\mu_{c}-\left(\omega +i/\tau \right)\hbar}{2\mu_{c}+\left(\omega +i/\tau \right)\hbar}\right),
\end{align}
\end{subequations}
where $\sigma_{0}=\pi e^2/\left(2h\right)$. The frequency dependences for the real and imaginary parts of the conductivity are shown in Figure~\ref{fig:2}a. Note that for terahertz, far- and mid-infrared (IR) radiations, the effect of the interband conductivity can be neglected~\cite{erdogan} for graphene with $\mu_{c}>0.2 \; \textrm{eV}$.

The basic properties of single quantum dots synthesized from $\textrm{Ag}_{2}\textrm{Se}$ QDs are presented in~\cite{Sahu,Park}. An important feature of $\textrm{Ag}_{2}\textrm{Se}$ QDs is a very small band gap, which makes them attractive quantum objects for controlling the photoluminescence spectrum by changing their radius~\cite{Kitai}. Such a dependence can be approximately described as follows
\begin{align}
\label{eq:4}
\omega_{12}&=\frac{eE_{g}}{\hbar}+\frac{2\hbar \kappa^{2}_{1,0}}{D^{2}_{QD}}\left(\frac{1}{m_{c}}+\frac{1}{m_{h}}\right)-\frac{3.56e^{2}}{\hbar\varepsilon_{st}D_{QD}4\pi\varepsilon_{0}},
\end{align}
where $m_{c}=0.32m_{0}$ and $m_{h}=0.54m_{0}$ are the effective masses of the electron and the hole~\cite{Sahu,Junod}, respectively; $E_{g}=0.07 \; \textrm{eV}$ is the bulk band gap of Ag2Se assigned to the tetragonal phase~\cite{Sahu,Abdullayev,Dalven,Baer}, $D_{QD}$ is the diameter of $\textrm{Ag}_{2}\textrm{Se}$ QD, $\varepsilon_{st}$ is the static dielectric constant of graphene (we assume $\varepsilon_{st}=11$~\cite{Gorbachev}), $\kappa_{1,0}=\pi$ is the root of the Bessel function, $m_{0}$ is the free electron mass, $\varepsilon_{0}$ is the electric constant.

Figure~\ref{fig:2}b shows the wavelength dependence of the interband $1S\left(e\right)--1S\left(h\right)$ transition on $\textrm{Ag}_{2}\textrm{Se}$ QD radius in MWIR. We assume that the QDs are pumped by a plane wave at a wavelength $\lambda=4.5 \; \textrm{\textmu m}$, which is normal to the graphene waveguide plane. Then, the relaxation of the excited QDs with a diameter $D_{QD}=5.79 \; \textrm{nm}$ in the proximity of the graphene sheet with a chemical potential $\mu_{c2}=0.3 \; \textrm{eV}$ leads to the effective excitation of the graphene SPPs. It is possible because the wavelength of the interband transition for such QDs is $\lambda_{0}=5.86 \; \textrm{\textmu m}$, see Figure~\ref{fig:2}b, and then the real part of the graphene permittivity takes a negative value, i.e., $\varepsilon_{gr}^{\textrm{R}}=1-\frac{\sigma^{\textrm{I}}}{\omega \Delta_{g}\varepsilon_{0}}<0$ (see Figure~\ref{fig:2}a), and the graphene behaves like a thin metal plate capable of supporting the SPPs at such a wavelength. On the contrary, the same QD nearby the graphene with $\mu_{c1}=0.1 \; \textrm{eV}$ does not lead to the excitation of the SPPs because $\varepsilon_{gr}^{\textrm{R}}>0$ at $\lambda_{0}$, and such graphene behaves like a dielectric. Note that the SPPs in a single graphene sheet are weak, and it is necessary to use a two graphene sheets waveguide arrangement~\cite{ourNanomaterials} to increase the power of surface waves. This requires additional tuning of the entire system.

Now we assume that the QDs are loaded into the center of the two graphene sheets plasmonic waveguide~\cite{Kim2} and used to pump the SPPs, as shown in Figure~\ref{fig:1}. At the same time, we consider the QDs as small defects that slightly change the kinetics of the propagating SPPs. Therefore, the dispersion equation for the SPP propagating in such a QD-loaded two graphene sheets waveguide can be written in the form~\cite{teng}
\begin{equation}
\label{eq:5}
-k_{h}\left(\pm e^{-k_{h}d}-1\right)=2ik_{0}c\varepsilon_{d}\varepsilon_{0}/\sigma,
\end{equation}
where $k_{h}=\sqrt{\beta^{2}-k^{2}_{0}}$, $\beta$ is the propagation constant of the SPP, $d$ is the distance between the sheets, $k_{0}=\frac{2\pi}{\lambda}$ is the wave vector of the incident electromagnetic field at wavelength $\lambda$ in a vacuum, $\varepsilon_{d}=1.619$ is the permittivity of the $\textrm{SiO}_{2}$ at $\lambda_{0}=5.864 \; \textrm{\textmu m}$ as a dielectric material surrounding the waveguide.

\begin{figure}[t]
\centering
\includegraphics[width=\columnwidth]{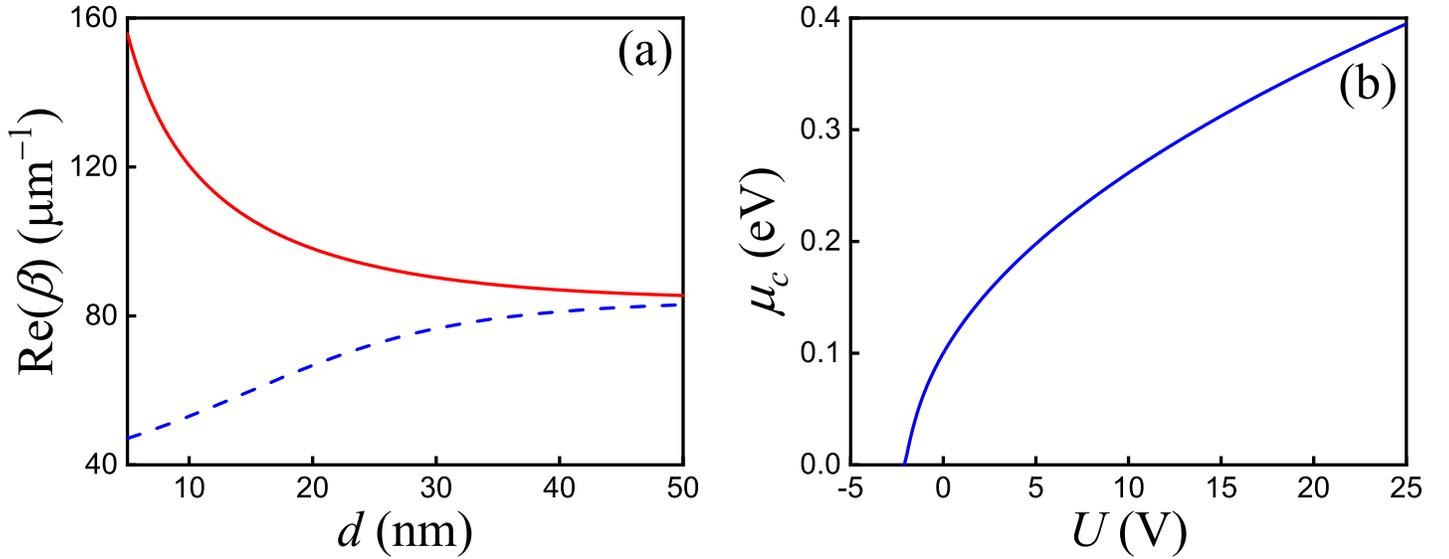}
\caption{\label{fig:3} (a) Propagation constants for the symmetric mode $\beta_{+}$ (solid red line) and the antisymmetric mode $\beta_{-}$ (dashed blue line) for the signal SPPs in the waveguide versus the distance between the graphene sheets. (b) The chemical potential dependence of graphene on the applied voltage (thickness of $\textrm{SiO}_{2}$ dielectric is $50 \; \textrm{nm}$, the no-voltage chemical potential is $\mu_{c0}=0.1 \; \textrm{eV}$, $T=300 \; \textrm{K}$).}
\end{figure}

Two dispersion branches for the two graphene sheets SPPs corresponding to the QD resonance wavelength $\lambda_{0}=5.864 \; \textrm{\textmu m}$ are presented in Figure~\ref{fig:3}a. Here, the red (blue) curve corresponds to the symmetric $\beta_{+}$ (antisymmetric $\beta_{-}$) mode~\cite{ali}, but we use only the symmetric mode $\beta_{+}$ since it concentrates as much energy as possible into the space between the sheets of the plasmonic waveguide. We choose $d=14 \; \textrm{nm}$ for the strong coupling regime of the SPP in graphene. This regime satisfies the condition $d<\xi$~\cite{wang1} because, in our case, $\xi=\textrm{Re}\left(\frac{\sigma}{ic\varepsilon_{0}\varepsilon_{d}k_{0}}\right)=23.71 \; \textrm{nm}$. The wavelength of the SPPs generated in a graphene waveguide is $\lambda_{SPP}=\frac{2\pi}{\textrm{Re}\left(\beta_{+}\right)}=58 \; \textrm{nm}$, and the propagation length is $L_{SPP}=\frac{\lambda_{0}}{4\pi \textrm{Im}\left(n_{EF}\right)}=250 \; \textrm{nm}$, where $n_{EF}=\frac{\beta_{+}}{k_{0}}=101$ is the effective refractive index of the SPP. These values are in good agreement with the electromagnetic simulation results based on the finite difference time domain (FDTD) method~\cite{Sullivan}, which we have realized by ourselves.

To control the kinetics of the SPPs generated in a plasmonic waveguide loaded with QDs, it is sufficient to change the electron density of the states either on one or both sheets. This can be done by applying the voltage independently to each graphene sheet~\cite{Gomez,Zheng} to change their chemical potential. At the same time, a set of tasks has to be solved for this implementation. In particular, if we use graphene plates on a substrate rather than single flakes, then the electron density distribution will initially be quite unstable and inhomogeneous. A good solution is to use hexagonal boron-nitride (hBN) as a buffer layer that significantly increases the homogeneity of the electron density in graphene, see Figure~\ref{fig:1}. In addition, high values of the field strength of the dielectric breakdown of the material make it possible to up the control voltage and vary the graphene chemical potential in a broader range~\cite{n13}.

As a result, the chemical potential dependence on the applied voltage is shown in Figure~\ref{fig:3}c, and taking into account the thickness of the $\textrm{SiO}_{2}$ layer between hBN and Au ($50 \; \textrm{nm}$), the voltage $13.7 \; \textrm{V}$ is sufficient to increase the initial chemical potential of graphene from $0.1 \; \textrm{eV}$ up to $0.3 \; \textrm{eV}$. The main point is that, under the chosen conditions, a two parallel graphene sheets plasmonic waveguide does not support the SPP modes if no external voltage is applied. In contrast, the SPPs are formed in accordance with Equation~(\ref{eq:5}) with an applied voltage of $0.3 \; \textrm{eV}$. Then, when the circuit in Figure~\ref{fig:1} is excited by a normally incident planar wave at a wavelength $4.5 \; \textrm{\textmu m}$~\cite{Qu}, the SPPs will be generated only in the area under the gold stripes with the applied voltage.

\section{Results. Voltage control of collective near-field effects in plasmonic \linebreak gold/graphene waveguides loaded with Ag2Se quantum dots}
The amplification and concentration of the near-field energy localized in a plasmonic waveguide can be realized by using an array of QDs and the synchronization of their responses. Such synchronization is achieved by varying the QD array period in the waveguide and tuning the voltage applied to graphene. First, we select the period $P$ of the QD array so that it is a multiple of the wavelength of the generated SPP, i.e., $P=n\lambda_{SPP}$, $n=1,2,3...$.

\begin{figure}[t]
\centering
\includegraphics[width=\columnwidth]{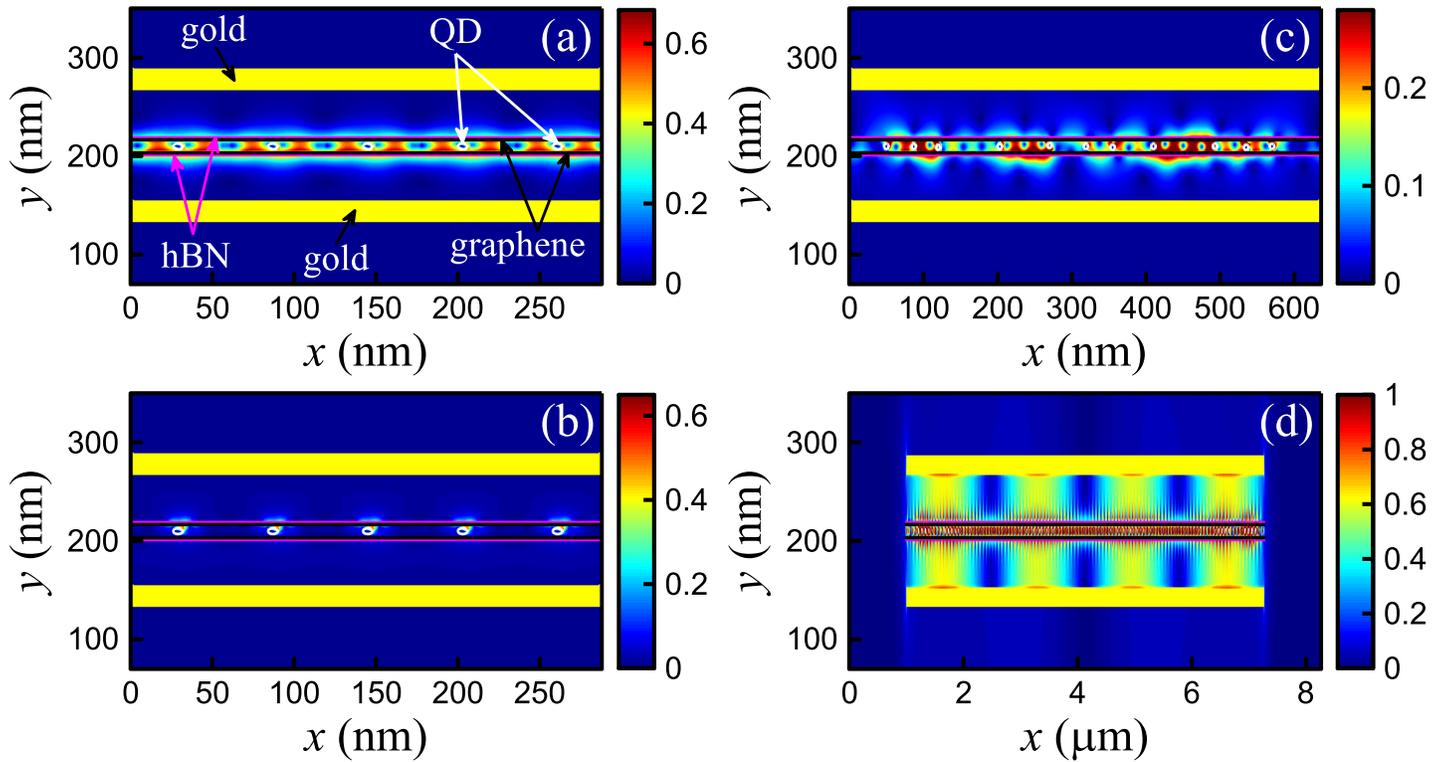}
\caption{\label{fig:4} Summarized electric field distributions (arbitrary unit) for a fragment of an infinite waveguide loaded with $5.79 \; \textrm{nm}$ $\textrm{Ag}_{2}\textrm{Se}$ QDs and the distance between the graphene sheets $14 \; \textrm{nm}$ for (a) the first order constructive interference for SPPs via an applied voltage $U_{1}=U_{2}=13.7 \; \textrm{V}$ ($\mu_{c}=0.3 \; \textrm{eV}$ for both sheets), (b) the near-field pattern for the no-voltage case ($\mu_{c}=0.1 \; \textrm{eV}$ for both sheets), (c) the near-field pattern for random QDs distribution and $U_{1}=U_{2}=13.7 \; \textrm{V}$. (d) The SPP-SPP modulation in the finite plasmonic gold/graphene waveguide with the constructive interference for the graphene SPPs ($U_{1}=U_{2}=13.7 \; \textrm{V}$). The thin magenta lines and the yellow bands depict the hBN layers and the gold stripes, respectively. The black lines correspond to graphene; the solid white lines depict the circle shape of the QDs.}
\end{figure}

Figure~\ref{fig:4}a shows the regime corresponding to the first order constructive interference for the SPP with $\lambda_{SPP}=58 \; \textrm{nm}$ in the plasmonic waveguide in accordance with Equation~(\ref{eq:5}). Here, the external voltage $U=13.7 \; \textrm{V}$ is applied to each graphene sheet to achieve the chemical potential value $\mu_{c2}=0.3 \; \textrm{eV}$. In this case, the strong collective response of the QDs creates a powerful SPP mode, which is localized in the space between the graphene sheets. Switching off the voltage on both sheets in Figure~\ref{fig:4}b returns the chemical potential of graphene to the value $\mu_{c1}=0.1 \; \textrm{eV}$, which results in a sharp decrease in the near-field intensity inside the waveguide due to the violation of the SPP generation conditions and the constructive interference for the QD array. The deposition of the metal stripes on the silica substrate in the required places allows routing of the SPPs generated by the excited QDs. At the same time, the other QDs located inside the plasmonic waveguide and outside the metal stripes are also excited by an external wave, but their energy is dissipated without the SPP generation since here $\hbar \omega > 2\mu_{c}$. From a technical perspective, the random distribution of the QDs in a waveguide is simpler to implement, more robust, and amenable to practical applications. However, the generated SPPs in such a waveguide with an applied voltage of $13.7 \; \textrm{V}$ behave unstably, making them difficult to use for near-field concentration and routing, see Figure~\ref{fig:4}c.

We choose such a topology in Figure~\ref{fig:4}d that the QDs can be used as sources of the SPPs generated in the space between the graphene sheets and the control gold stripes. In the latter case, the SPP wavelength in the gold stripes waveguide is $3300 \; \textrm{nm}$, which can be calculated using Equation~(\ref{eq:5}). This wavelength corresponds to the SPP-SPP modulation, which should be attributed to the interaction and the near-field energy repumping between the graphene-localized SPPs and the gold-localized SPPs. At the same time, we do not consider the change in the electron density on the live gold contacts, but this correction will be negligible. As a result, the internal graphene waveguide acts as an efficient near-field pumping for the external gold waveguide. Note that the injection of defects by removing one or several QDs from the circuit makes it possible to change the field distribution to the required localization in the waveguide. This effect can be used to pump nanostructures~\cite{our3}, particularly the array of nanoantennas integrated or connected to a plasmonic waveguide. Additional control of the external voltage allows switching on/off and changing the emission regimes of an individual nanoantenna and varying the radiation pattern of all arrays, in general.

\section{Conclusion}
In this paper, we considered the graphene-based waveguide model loaded with the QDs whose electromagnetic properties are controlled by applying an external voltage. We optimized the parameters of two parallel graphene sheets and QDs to realize the controllable excitation of SPPs in the graphene waveguide. It was shown that as the control parameter of the system can be adopted, the magnitude of the applied voltages modifies the graphene chemical potential and leads to a change in the regime of the SPPs excitation. The controllable SPP excitation results in a high concentration of the near-field energy and its routing along the local paths determined by the spatial configuration of the applied external voltage.

In general, the presented system of the graphene waveguide loaded with QDs and the voltage control can be used to design ultra-compact optical devices for transport of the near-field energy, its concentration, and pumping of light-emitting nanostructures.




\medskip
\textbf{Acknowledgements} \par 
This work was supported by the Russian Science Foundation, Grant No. 20-12-00343. The numerical algorithm development for the simulation of SPPs in 2D materials with QDs and the analysis of the calculated results was supported by the Ministry of Science and Higher Education of the Russian Federation within the state task VlSU (GB 1187/20). The authors thank to MIPT Language Training and Testing Center (LTTC) for the help with English language editing.

\medskip

%
\bibliographystyle{MSP}
\bibliography{Arxiv}

\end{document}